\documentclass[useAMS,usenatbib]{mn2e}
\usepackage{graphicx}
\usepackage{amsmath}
\usepackage{amssymb}

\def\dark{1}
\def\mpe{2}
\def\roe{3}
\def\iaa{4}
\def\eso{5}
\def\tls{6}
\def\puc{7}
\def\mil{8}
\def\wei{9}

\title[Revised Host Association for GRB 020819B]{A Revised Host Galaxy Association for GRB\,020819B: A High-Redshift Dusty Starburst, Not a Low-Redshift Gas-Poor Spiral}

\author[D. A. Perley et al.]
{Daniel A. Perley$^{\dark}$\thanks{e-mail: dperley@dark-cosmology.dk}, 
Thomas Kr\"uhler$^{\mpe}$,
Patricia Schady$^{\mpe}$, 
\newauthor 
Micha{\l} J.~Micha{\l}owski$^{\roe}$,
Christina C. Th\"one$^{\iaa}$,
Dirk Petry$^{\eso}$,
John F. Graham$^{\mpe}$,
\newauthor 
Jochen Greiner$^{\mpe}$,
Sylvio Klose$^{\tls}$,
Steve Schulze$^{\puc,\mil,\wei}$, and
Sam Kim$^{\puc}$
\\
\\
$^{\dark}${Dark Cosmology Centre, Niels Bohr Institute, University of Copenhagen, Juliane Maries Vej 30, 2100 K{\o}benhavn {\O}, Denmark} \\
$^{\mpe}${Max-Planck-Institut f\"ur extraterrestrische Physik, 85740 Garching, Giessenbachstr. 1, Germany} \\
$^{\roe}${Scottish Universities Physics Alliance, Institute for Astronomy, University of Edinburgh, Royal Observatory, Edinburgh, EH9 3hJ, UK} \\
$^{\iaa}${Instituto de Astrof\'isica de Andaluc\'ia, Glorieta de la Astronom\'ia s/n, E-18008 Granada, Spain} \\
$^{\eso}${European Southern Observatory, Karl-Schwarzschild-Strasse 2, 85748 Garching, Germany} \\
$^{\tls}${Th\"uringer Landessternwarte Tautenburg, Sternwarte 5, D-07778 Tautenburg, Germany} \\
$^{\puc}${Instituto de Astrof\'isica, Pontificia Universidad Cat\'olica de Chile, Vicu\~na Mackenna 4860, 7820436 Macul, Santiago, Chile} \\ 
$^{\mil}${Millennium Institute of Astrophysics, Vicu\~na Mackenna 4860, 7820436 Macul, Santiago, Chile} \\
$^{\wei}${Department of Particle Physics and Astrophysics, Weizmann Institute of Science, Rehovot 76100, Israel} \\
}

\begin{document}

\date{}

\pagerange{\pageref{firstpage}--\pageref{lastpage}} \pubyear{2016}

\def\LaTeX{L\kern-.36em\raise.3ex\hbox{a}\kern-.15em
    T\kern-.1667em\lower.7ex\hbox{E}\kern-.125emX}

\newcommand{\citepeg}[1]{\citep[{e.g.,}][]{#1}}
\newcommand{\citepcf}[1]{\citep[{see}\phantom{}][]{#1}}
\newcommand{\rha}[0]{\rightarrow}
\newcommand\ion[2]{\text{#1\,\textsc{\lowercase{#2}}}}	
\def\etal{{\sl et al.}}
\def\lsim{\hbox{ \rlap{\raise 0.425ex\hbox{$<$}}\lower 0.65ex\hbox{$\sim$}}}
\def\gsim{\hbox{ \rlap{\raise 0.425ex\hbox{$>$}}\lower 0.65ex\hbox{$\sim$}}}
\def\arcmin{\hbox{$^\prime$}}
\def\arcsec{\hbox{$^{\prime\prime}$}}
\def\arcdeg{\mbox{$^\circ$}}
\def\fd{\hbox{$~\!\!^{\rm d}$}}
\def\fh{\hbox{$~\!\!^{\rm h}$}}
\def\fm{\hbox{$~\!\!^{\rm m}$}}
\def\fs{\hbox{$~\!\!^{\rm s}$}}
\def\ale{\mathrel{\hbox{\rlap{\hbox{\lower4pt\hbox{$\sim$}}}\hbox{$<$}}}}
\def\age{\mathrel{\hbox{\rlap{\hbox{\lower4pt\hbox{$\sim$}}}\hbox{$>$}}}}
\def\msyr{\hbox{M$_\odot$ yr$^{-1}$}}
\def\oii{\ion{O}{II}}
\def\oiii{\ion{O}{III}}
\def\nii{\ion{N}{II}}
\def\Swift{{\textit{Swift}}\,}
\def\Fermi{{\textit{Fermi}}\,}
\def\apjl{{ApJL}\,}
\def\apj{{ApJ}\,}
\def\nat{{Nature}\,}
\def\aap{{A\&A}\,}
\def\pasp{{PASP}\,}
\def\aj{{AJ}\,}
\def\procspie{{Proc.~SPIE}\,}
\def\mnras{{MNRAS}\,}
\def\apjs{{ApJS}\,}
\def\aaps{{A\&AS}\,}
\def\araa{{ARAA}\,}
\def\apss{{Ap\&SS}\,}
\def\physrep{{PhysRep}\,}
\def\href{{}}

\label{firstpage}
\label{lastpage}

\maketitle

\begin{abstract}
The purported spiral host galaxy of GRB\,020819B at $z=0.41$ has been seminal in establishing our view of the diversity of long-duration gamma-ray burst environments: optical spectroscopy of this host provided evidence that GRBs can form even at high metallicities, while millimetric observations suggested that GRBs may preferentially form in regions with minimal molecular gas.   We report new observations from VLT (MUSE and X-shooter) which demonstrate that the purported host is an unrelated foreground galaxy. The probable radio afterglow is coincident with a compact, highly star-forming, dusty galaxy at $z=1.9621$.  The revised redshift naturally explains the apparent nondetection of CO(3-2) line emission at the afterglow site from ALMA.  There is no evidence that molecular gas properties in GRB host galaxies are unusual, and limited evidence that GRBs can form readily at super-Solar metallicity.\end{abstract}


\section{Introduction}
\label{sec:intro}

The majority of long-duration gamma-ray bursts (GRBs) in the low-redshift universe originate from low-mass, low-metallicity, irregular dwarf galaxies \citepeg{LeFloch+2003,Fruchter+2006,Modjaz+2008,Graham+2013,Perley+2016}.  GRB\,020819B\footnote{Occasionally designated as simply GRB\,020819.} stands out as the most notable exception.  This burst was detected by the High Energy Transient Explorer (HETE-2) two years prior to the launch of Swift; follow-up using the Very Large Array identified a fading radio source (the probable afterglow) positionally coincident with the outskirts of a large spiral galaxy at $z=0.41$ (\citealt{Jakobsson+2005}; J05).  The spiral galaxy is massive \citep{KupcuYoldas+2010}, and metal-rich (\citealt{Levesque+2010}; L10).  Although the location of the radio afterglow is far (3$\arcsec$, or 16 kpc in projection) from the spiral's nucleus and a ``blob'' of optical emission distinct from the rest of the galaxy is visible at this location, spectroscopy reported by L10 established that this blob was at the same redshift as the spiral and that its metallicity was above Solar (12+log[O/H] = $9.0\pm0.1$, on the \citealt{KD02} diagnostic).  This represented perhaps the clearest demonstration that GRBs can successfully form even in metal-rich environments.

\begin{figure}
\centerline{
\includegraphics[width=3.3in,angle=0]{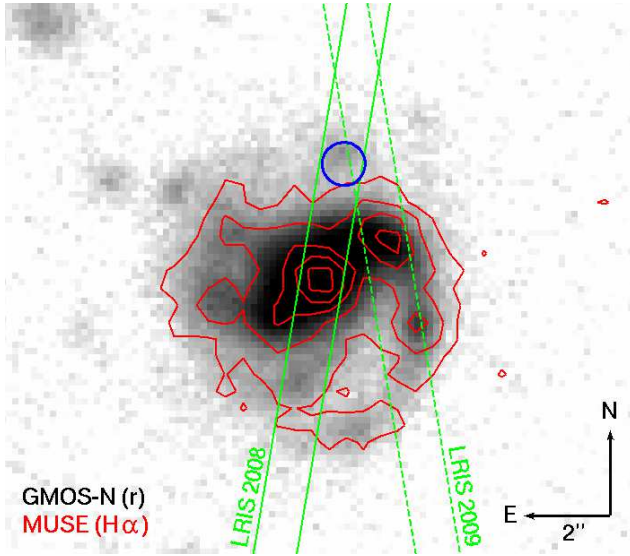}} 
\caption{Optical (GMOS $r$-band; grayscale) and H$\alpha$ (MUSE; contours) images of the region around GRB\,020819B.  The radio afterglow position is marked as a blue circle (uncertainty radius 0.5$\arcsec$) and is consistent with the location of a marginally-extended ``blob'' of optical emission.  Green lines indicate the inferred slit positions for the observations of L10 (solid: 2008, dashed: 2009).  No significant H$\alpha$ emission at $z=0.41$ is present at the afterglow site in the MUSE image, or in the LRIS or X-shooter spectra.}
\label{fig:imaging}
\end{figure}

The properties of this galaxy have also attracted follow-up at longer wavelengths.  It is one of the few GRB hosts to be observed with the Atacama Large Millimetre Observatory (ALMA) to date, as part of the study of \citealt{Hatsukade+2014} (H14).  While a source at the afterglow site is well-detected in the millimetre continuum ($F_\nu = 0.14$ mJy at 1.2\,mm), indicating that this region is rich in dust, \emph{no} molecular gas emission was detected in spectroscopic-mode observations covering the expected wavelength of the CO(3-2) emission line at $z=0.41$.  The inferred gas-to-dust ratio is lower than for any well-studied class of galaxy, suggesting that GRBs may prefer regions of very low molecular gas density.  This could originate from a different mode of star-formation in these regions, possibly involving the direct collapse of atomic gas \citep{Michalowski+2015}. 

No optical counterpart was identified for this burst and therefore no absorption redshift was obtained, so its association with the massive spiral hinges on two important assumptions.  First, the variable radio source reported in the error circle by J05 must indeed be the GRB afterglow.  Second, this radio source must have originated from within the spiral galaxy.  These two assumptions are justified primarily by statistical arguments.  The probability of an unrelated fading radio source appearing in the HETE-2 X-ray error circle and the probability of a source aligning with an unrelated massive foreground galaxy are both low ($\sim$\,$10^{-2}$), but not so low as to be implausible.  

The considerable importance given to this GRB merits further scrutiny of this issue.  In this Letter, we re-analyze the archival data originally presented by L10 and H14 and also present extensive new observations of this system, including near-infrared spectroscopy of the blob at the afterglow site.  Our observations demonstrate that the reported GRB counterpart lies in a background galaxy, unrelated to the $z=0.41$ spiral widely assumed by previous authors to be the host.  
Our observations and basic results are presented in \S 2.  We reassess the probabilistic arguments used to associate the GRB with the radio counterpart, and the radio counterpart with the candidate host galaxies, in \S 3.  Our conclusions, including a summary of the properties of the high-redshift galaxy that represents the probable host, are then presented in \S 4.

\section{Observations and Results}
\label{sec:observations}

\subsection{MUSE observations}
\label{sec:muse}

The putative host galaxy was observed using the Multi Unit Spectroscopic Explorer (MUSE; \citealt{Bacon+2010}) integral field unit spectrograph at the VLT.  Observations were obtained on 2016-07-09 as part of ESO program 097.D-1054 and consisted of three dithered exposures with an integration time of 1200~s each. They cover a field of view of 1$\arcmin\times$1$\arcmin$ sampled in spatial pixels of 0.2$\arcsec$ and in wavelength steps of 1.25\,\AA\ spanning a wavelength range of 4800--9300\,\AA. 

The MUSE data were reduced and calibrated in a standard manner using version \texttt{1.6} of the ESO pipeline \citep{Weilbacher+2014}. We cleaned sky-background residuals that were present after the pipeline reduction as described in \citet{Soto+2016}. The stellar PSF has a full-width at half maximum of 0.9$\arcsec$ in a reconstructed MUSE image centred around 9000\,\AA.  A small astrometric shift was then applied to align the final data cube with archival Gemini-North Multi-Object Spectrograph (GMOS) imaging of the field.

An H$\alpha$ map of the galaxy formed from the data cube is presented in Figure \ref{fig:imaging} (contours).  The radio afterglow location is designated by a circle.  We do not detect any significant H$\alpha$ emission at this location to a limiting flux of $F_{{\rm H}\alpha}<0.8 \times 10^{-17}$ erg cm$^{-2}$s$^{-1}$.  We also extracted a one-dimensional spectrum at the location of the radio transient and detected no lines anywhere within our spectral range, which covers all major emission lines at $z=0.41$.

\begin{figure*}
\centerline{
\includegraphics[width=6.5in,angle=0]{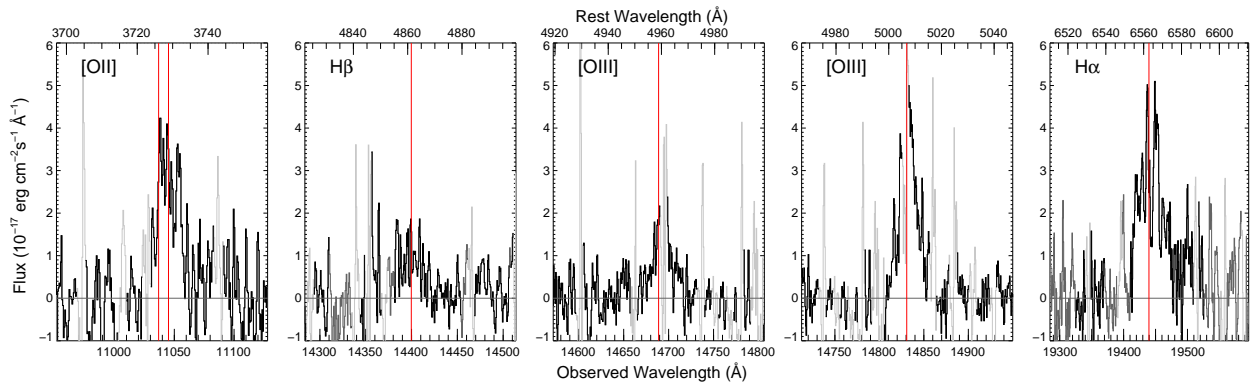}} 
\caption{X-shooter observations reveal the redshift of the likely host of GRB\,020819B (the ``blob'' at the radio position in Figure 1) to be $z=1.9621$.  We detect strong emission lines from [\oii]$\lambda$3726,3729, [\oiii]$\lambda$5007, and H$\alpha$, as well as weaker emission from [\oiii]$\lambda$4959 and H$\beta$.  Red lines show central wavelengths of the individual transitions at this redshift.  Regions of the spectrum strongly affected by telluric absorption or sky-line emission residuals are shown in a lighter colour.}
\label{fig:xshooter}
\end{figure*}

\subsection{Archival Keck observations}
\label{sec:keck}

The MUSE data directly contradict the reported strong H$\alpha$ and [\nii] detections at the GRB location reported by L10 using LRIS observations acquired in November 2009 ($F_{{\rm H}\alpha}=41.4\times10^{-17}$~erg\,cm$^{-2}$s$^{-1}$).  We downloaded their observations from the Keck Observatory Archive and independently reanalyzed them using our own data analysis tools.  Although L10 report that they chose a position angle (PA) to cover the host nucleus and afterglow site simultaneously, at the PA value of 9.5$\arcdeg$ quoted in their paper (confirmed by archive metadata) this would not have been possible.  Comparison of the two-dimensional spectra to our continuum and H$\alpha$ imaging demonstrates that the slit was positioned west of the nucleus, roughly at the position indicated by the dashed lines in Figure \ref{fig:imaging}.  The two traces visible in the 2D spectrum (interpreted by L10 as the ``nucleus'' and ``site'' spectra) correspond to star-forming regions in the western spiral arm.  The slit partially intersected the blob at the afterglow site, but no line emission is detected at this position in the 2D spectra.


L10 also observed the host system in November 2008 at a PA of 345.88$\arcdeg$.  These observations do unambiguously cover both the nucleus and the GRB location (solid green lines in Figure \ref{fig:imaging}).  We downloaded and reanalyzed these observations also; no significant line emission is observed at the location of the blob in the 2D spectrum.

\subsection{X-shooter observations}
\label{sec:xshooter}

The nondetection of emission lines at both optical and millimeter wavelengths at the afterglow site led us to suspect that the ``blob'' at this location, treated as a star-forming region within the $z=0.41$ galaxy by L10, H14, and other authors, is a separate galaxy at significantly higher redshift.  To test this hypothesis, we acquired optical and NIR spectroscopy of the source using X-shooter \citep{Vernet+2011} on the VLT via ESO program DDT 297.A-5055.  

Data were obtained on two different nights starting on 2016-08-08 and 2016-08-30.  Observations consisted of several nodded frames with a total on-source integration time of 4960\,s in the optical arms (wavelength range 3200--10000\,\AA) and 5280\,s in the NIR arm (10000--25000\,\AA).  X-shooter's slits were positioned using blind offsets from a field star, and oriented at the parallactic angle.  To reduce, combine and calibrate our X-shooter data, we used the ESO-supplied pipeline \citep{Goldoni+2006} and our own software.  

Several emission lines are securely detected in the NIR portion of the reduced spectrum (Figure \ref{fig:xshooter}) at the afterglow location.  Particularly strong ($>$10$\sigma$) lines are observed at (air, heliocentric) wavelengths of 14830 and 19440 \AA.  These wavelengths identify them unambiguously as [\oiii]$\lambda$5007 and H$\alpha$ (respectively) at a common redshift of $z = 1.9621\pm0.0001$.  We also detect, at lower significance (3--6$\sigma$), emission lines at the wavelengths of redshifted H$\beta$, [\oiii]$\lambda$4959 and [\oii]$\lambda$3727.  We do not detect significant emission from [\nii] or Lyman-$\alpha$, or at any other wavelength (although optical emission lines from the $z=0.41$ galaxy can be seen elsewhere on the slit). 

The velocity width of both strong lines is large at $\sigma=190\pm20$ km\,s$^{-1}$, but much less than broad-line AGN.  Likewise, the line ratios between [\oiii], H$\beta$, [\nii], and H$\alpha$ are characteristic of star-forming galaxies \citep{Baldwin+1981} and rule out significant contribution from an active nucleus.  Assuming the line flux to originate entirely from star-formation, the H$\alpha$ flux corresponds to a minimum star-formation rate (SFR) of 21 M$_\odot$yr$^{-1}$ (\citealt{Kennicutt+1998}; we assume a cosmology of $\Omega_M$=0.3, $\Omega_\Lambda$=0.7, $h$=0.7).  Incorporating a dust-extinction correction of $E_{B-V} = 0.55_{-0.27}^{+0.37}$ mag from the Balmer decrement, this increases to $76^{+103}_{-35} M_\odot$yr$^{-1}$.  

The weak detection of H$\beta$ and nondetection of [\nii] make it difficult to estimate metallicity or other line parameters precisely.  We ran the Monte Carlo code of \cite{Bianco+2016} on our line flux values (Table \ref{tab:hostproperties}) to calculate the metallicity using various diagnostics and calibrations, and estimate a value between 8.29 (using \citealt{PP04} O3N2) and 8.81 (using \citealt{KK04} N2H$\alpha$), corresponding to approximately 0.5--1.1 $Z_\odot$.

\subsection{ALMA Reanalysis}
\label{sec:alma}

The ALMA observations first reported by H14\footnote{Program ID JAO.ALMA\#2011.0.00232.S.  ALMA is a partnership of ESO, NSF, NINS, NRC, NSC, ASIAA, and KASI, in cooperation with the Republic of Chile. The Joint ALMA Observatory is operated by ESO, AUI/NRAO and NAOJ.}, covering four 1.875 GHz spectral frequency windows centred at 245.072, 246.947, 260.500 and 262.375 GHz, would not have covered the frequencies of any strong molecular transitions at $z=1.9621$.   Nevertheless, we carefully reanalyzed the observations in the standard ALMA data analysis package CASA (version 4.6), performing a blind search of the entire spectrum at the afterglow site for line features.  None were found, although we verify the continuum detection at a flux level consistent with value originally reported by H14.

\section{Discussion}
\label{sec:discussion}

Our results rule out the conclusions of L10 and H14 which claim that the local environment of this GRB has highly super-Solar metallicity or that it is depleted in molecular gas.  No strong constraints can currently be placed on either the metallicity or the molecular gas mass in the higher-redshift galaxy at the GRB position with our data, although the metallicity is probably subsolar.

The galaxy at $z=1.96$ is a remarkable object in its own right, and may be relevant to understanding GRB progenitors \emph{if} indeed this source is genuinely the host of GRB\,020819B.  This association is not definitive, although we argue that it is likely.  The two alternative possibilities (that the radio counterpart is not the afterglow, or that the radio counterpart is genuine but that $z=1.96$ galaxy is not the host) are addressed below.

\subsection{Is the purported radio afterglow of GRB020819B associated with the burst?}

As we have noted already, the afterglow of GRB\,020819B was localized only in the radio band; furthermore the observations are only at one frequency and cover a limited span in time.  It is at least possible that the source represents an unrelated variable object, such as an AGN or quasar.  
Fortunately, the final HETE-2 error circle is relatively small ($\sim$80$\arcsec$ in radius; \citealt{Villasenor+2004}), so the probability of coincidental appearance of a variable source is not high.   \cite{GCN1842} estimated the probability of finding a strongly-variable source in the original HETE-2 error circle to be 2\%; our own independent calculation using variability statistics from \cite{Ofek+2011} provides a similar value of 4\%.  
Other lines of argument also support association with the GRB: the afterglow fades monotonically and exhibits no further flaring activity over observations spanning $>$100 days, and it is localized to a star-forming, high-redshift galaxy with no direct evidence of AGN activity.

It is still conceivable that the source is not really the afterglow.  It is interesting, in particular, that radio emission remains detectable at this location more than 10 years after the burst (\citealt{Greiner+2016} report a flux density of $31\pm8$ $\mu$Jy at 3 GHz).  This is more than can be readily explained by star-formation: it would imply an SFR \citep{Murphy+2011} of $\sim$800\,$M_\odot$\,yr$^{-1}$, conflicting with the estimate from the ALMA continuum.  It could represent lingering emission from the GRB afterglow, but this would require a much slower late-time decay than predicted by standard models \citep{Greiner+2016}.  Further observations are needed: the discovery of additional flares would greatly decrease our confidence in the association and suggest that the GRB may have had nothing to do with either of the two galaxies at this location.  In the meantime, however, it is reasonable to proceed assuming that the radio counterpart identified in J05 does indeed represent the radio afterglow.

\subsection{Is the radio counterpart associated with the high-redshift coincident galaxy?}

We can also question whether it is actually the high-redshift galaxy that represents the chance alignment: perhaps the GRB does indeed originate in the outskirts of the $z=0.41$ galaxy, but its location happened to line up with an unrelated, luminous galaxy at higher redshift!  The statistical probabilities of an arbitrary position on the sky landing within 3$\arcsec$ of an $R=19.5$ mag galaxy and within 0.5$\arcsec$ of an $R=24.0$ mag galaxy are in fact rather similar.  

Several lines of argument favor associating the GRB with the higher-redshift counterpart.  Most notably, there is extensive star-formation occurring within the high redshift galaxy and \emph{none} (down to our detection limits) in the section of the lower redshift galaxy consistent with the observed afterglow position.  GRBs are generally observed to explode in the densest and most active star-forming regions of their hosts, typically close to the centres \citep{Bloom+2002,Fruchter+2006,Blanchard+2016}; to find one in the outskirts of a galaxy in a region with no detectable star-formation or continuum emission to deep limits would be very unusual.   

Furthermore, this is a ``dark'' burst with no optical afterglow detection: reanalysis of the NIR data of \cite{Klose+2003} confirms no variability at the putative radio afterglow position or anywhere else in the final HETE-2 error circle (limiting magnitude $K>19$).  The dark burst host population is dominated by massive, highly star-forming, and dusty galaxies at $z=1-3$ \citepeg{Kruehler+2011,Rossi+2012,Perley+2013a}; the properties of the background galaxy are consistent with this population.  The foreground galaxy, while massive, contains little dust (H14) and no evidence for obscured star-formation outside its nucleus.

Even if the GRB did originate from the halo of the $z=0.41$ galaxy, the results of L10 and H14 would still be ruled out, since the metallicity and dust-to-gas ratio at the afterglow site cannot be constrained by their observations.

\section{Conclusions}
\label{sec:conclusions}

A brief summary of the available multiwavelength constraints of the new, probable host galaxy is provided in Table \ref{tab:hostproperties}.  
The H$\alpha$ luminosity and the millimetre flux (fit against a variety of starburst galaxy templates; \citealt{Silva+1998,Michalowski+2010}) both imply an SFR of between 30--120\,$M_\odot$\,yr$^{-1}$, consistent with a luminous infrared galaxy (LIRG).  LIRGs are uncommon among GRB hosts generally, but have been observed to host heavily obscured GRBs before \citepeg{LeFLoch+2006,Perley+2013b}.

Gravitational lensing by the foreground galaxy could conceivably contribute to the apparently high luminosity.  
The foreground galaxy's halo mass is unknown, but for $M_{\rm encl}=10^{12}$\,$M_\odot$ the Einstein radius would overlap the burst location, magnifying both the GRB and its host by a potentially large factor.  This would provide a natural explanation for the otherwise-coincidental foreground alignment and decrease the inferred luminosity and SFR of the high-$z$ galaxy somewhat.  Direct evidence for lensing in the form of multiple images or episodes is, however, lacking.

The metal dependence of GRB formation has been hotly debated for the past decade.  As the GRB host with the highest well-determined metallicity (both overall and site-specific), GRB\,020819B has been instrumental in establishing a case for a GRB progenitor that is able to function at metallicities above $Z_\odot$.  Our discovery that the nearby spiral is not the host therefore has important implications for the GRB metallicity dependence within the super-Solar regime.  While there are other cases of super-Solar metallicity hosts \citepeg{Elliott+2013,Kruehler+2015}, most of these originate at high redshifts where offsets and metallicity gradients are difficult to measure.  Low-$z$ examples are sparing, and generally also not secure: GRB\,050826, for example (12+log[O/H]=8.83; \citealt{Levesque+2010b}), also lacks an afterglow absorption spectrum or an associated supernova.  Swift has now detected over 1000 bursts, some of which ($\sim$1\%; \citealt{Cobb+2008}) are destined to closely align with foreground galaxies.  While it may yet be the case that GRBs can and do form above Solar metallicity, the frequency of such events is (at minimum) likely to be lower than previously believed.  We suggest that future studies of this topic approach low-$z$ host associations not confirmed via supernova or afterglow spectroscopy with increased caution.

More recently, the role of (deficient) molecular gas in GRB (progenitor) formation has also aroused significant interest and debate.  
Direct evidence for a molecular gas deficiency via millimetric spectral-line observations is extremely sparse, based almost entirely on the singular example of the GRB 020819B site reported by H14.  In fact, the other host galaxy observed by H14 (GRB\,051022) had a low but not anomalous gas-to-dust ratio; likewise, the gas mass in the host of GRB 080517 \citep{Stanway+2015} is normal, although it is being converted into stars very rapidly.  With the GRB\,020819B result explained as a redshift mismatch, there is no direct evidence that the large-scale molecular gas properties around GRB sites are unusual.

\begin{table}
\begin{minipage}{90mm}
\caption{Properties of the Probable Host Galaxy}
\label{tab:hostproperties}
\begin{tabular}{lll}
\hline
 Property   & Value & Reference \\
\hline
 Redshift                  & 1.9621 $\pm$ 0.0001                         &       \\
 $B$ mag                   & 26.10 $\pm$ 0.50 $^{a}$                     & J05 \\
 $R$ mag                   & 23.98 $\pm$ 0.06                            & J05 \\
 $K$ mag                   & 20.80 $\pm$ 0.30                            & J05 \\
 H$\alpha$ flux            & 14.9 $\pm$ 1.2 $^{b}$                       &    \\
 H$\beta$ flux             & 3.0 $\pm$ 0.9                               &    \\
 \oii$\lambda$3727 flux     & 7.2 $\pm$ 2.0                               &    \\
 \oiii$\lambda$4959 flux    & 3.2 $\pm$ 0.6                               &    \\
 \oiii$\lambda$5007 flux    & 9.7 $\pm$ 0.7                               &    \\
 \nii$\lambda$6583 flux     & 2.9 $\pm$ 1.7                               &    \\
 1.2 mm flux density       & 140 $\pm$ 30 $\mu$Jy                        & H14   \\
 3 GHz flux density        & 31 $\pm$ 8 $\mu$Jy                          & \citealt{Greiner+2016} \\
 Velocity width ($\sigma$) & 190$\pm$20 km\,s$^{-1}$                      &   \\
\hline
\end{tabular}
\\
$^{a}$ Magnitudes are Vega, uncorrected for foreground extinction. \\
$^{b}$ Units of all line fluxes are 10$^{-17}$ erg cm$^{-2}$s$^{-1}$ \\
\end{minipage}
\end{table}

\vskip 0.02cm

\section*{Acknowledgments}

DAP acknowledges support from a Marie Sk{\l}odowska-Curie Individual Fellowship within the Horizon 2020 EU Framework Programme for Research and Innovation (H2020-MSCA-IF-2014-660113).  PS and TK acknowledge support through the Sofja Kovalevskaja Award to PS.   SS and SK acknowledge support from FONDEYT (310534, 3130488).   This research has made use of the Keck Observatory Archive (KOA), which is operated by the W. M. Keck Observatory and the NASA Exoplanet Science Institute (NExScI), under contract with the NASA.  We thank E.~Levesque for confirming the orientation of the archival Keck observations.  We also thank D.~A.~Kann, M.~Modjaz, and the anonymous referee for useful comments, and acknowledge useful discussions with D.~Watson, J.~Fynbo, and J.~Hjorth.  

\bibliographystyle{apj}

\end{document}